\documentclass[twoside,11pt]{article}

\usepackage{jmlr2e}
\usepackage{mathrsfs}
\usepackage{natbib,graphicx,setspace,lscape,longtable,xr}
\usepackage{mathrsfs,amsmath,amsthm,amssymb,color}
\usepackage{natbib,epsfig,graphicx,pdfpages}
\usepackage{rotating}
\usepackage{float}
\usepackage{bm}
\usepackage{ulem}
\usepackage{CJK}
\usepackage{ragged2e}
\usepackage{setspace}
\usepackage{threeparttable}
\usepackage{multirow}
\usepackage{enumitem}
\usepackage{hyperref}
\usepackage{etoolbox}
\usepackage{subfig}
\usepackage[ruled]{algorithm2e}
\usepackage{adjustbox}
\usepackage{blindtext}
\usepackage{lineno}

\newtheorem{theorem}{Theorem}
\newtheorem{lemma}{Lemma}
\newtheorem{proposition}{Proposition}
\def\bet{\begin{theorem}}
\def\eet{\end{theorem}}
\def\bel{\begin{lemma}}
\def\eel{\end{lemma}}
\def\bep{\begin{proposition}}
\def\eep{\end{proposition}}

\def\vec{\mathrm{vec}}
\def\vech{\mathrm{vech}}

\def\beq{\begin{equation}}
\def\eeq{\end{equation}}
\def\beqr{\begin{eqnarray}}
\def\eeqr{\end{eqnarray}}
\def\beqrs{\begin{eqnarray*}}
\def\eeqrs{\end{eqnarray*}}
\def\bet{\begin{theorem}}
\def\eet{\end{theorem}}
\def\bel{\begin{lemma}}
\def\eel{\end{lemma}}
\def\bep{\begin{proposition}}
\def\eep{\end{proposition}}
\def\bg{\begin{figure}[tbph]\begin{center}}
\def\eg{\end{center}\end{figure}}

\def\bc{\begin{center}}
\def\ec{\end{center}}

\def\wt{\widetilde}

\def\wh{\widehat}

\def\mR{\mathbb{R}}

\def\mL{\mathcal L}

\def\mM{\mathcal M}
\def\mF{\mathcal F}

\def\mX{\mathbb{X}}

\def\mY{\mathbb{Y}}

\def\argmax{\mbox{argmax}}

\jmlrheading{}{2023}{1-32}{4/21}{10/00}{Li2022}{Li, Zhou and Wang}

\ShortHeadings{GMM with Rare Events}{Li, Zhou and Wang}
\firstpageno{1}

\begin{document}

\title{Gaussian Mixture Model with Rare Events}

\author{\name Xuetong Li \email 2001110929@stu.pku.edu.cn \\
\addr Guanghua School of Management\\
Peking University\\
Beijing, China
\AND
\name Jing Zhou \email   jing.zhou@ruc.edu.cn\\
\addr Center for Applied Statistics, School of Statistics \\
Renmin University of China\\
Beijing, China
\AND
\name Hansheng Wang \email hansheng@gsm.pku.edu.cn \\
\addr Guanghua School of Management\\
Peking University\\
Beijing, China}

\editor{ }

\maketitle

\begin{abstract}%   <- trailing '%' for backward compatibility of .sty file
We study here a Gaussian Mixture Model (GMM) with rare events data.
In this case, the commonly used Expectation-Maximization (EM) algorithm exhibits extremely slow numerical convergence rate.
To theoretically understand this phenomenon, we formulate the numerical convergence problem of the EM algorithm with rare events data as a problem about a contraction operator.
Theoretical analysis reveals that the spectral radius of the contraction operator in this case could be arbitrarily close to 1 asymptotically.
This theoretical finding explains the empirical slow numerical convergence of the EM algorithm with rare events data.
To overcome this challenge, a Mixed EM (MEM) algorithm is developed, which utilizes the information provided by partially labeled data.
As compared with the standard EM algorithm, the key feature of the MEM algorithm is that it requires additionally labeled data.
We find that MEM algorithm significantly improves the numerical convergence rate as compared with the standard EM algorithm.
The finite sample performance of the proposed method is illustrated by both simulation studies and a real-world dataset of Swedish traffic signs.
\end{abstract}

\begin{keywords}
Rare Events Data, Gaussian Mixture Model, Expectation-Maximum Algorithm, Partial Labeled Data, Unsupervised Learning
\end{keywords}

\section{Introduction}

In this study, we investigate a problem related to the analysis of rare events data.
Rare events data are defined as a type of data with a binary response ($Y\in\mR^{1}$) and feature vector ($X\in\mR^p$).
Furthermore, we require that the probability of the response belonging to one particular class (e.g., $Y=1$) is extremely small.
For convenience, we refer to this class (i.e., the class with $Y=1$) as the minor class, representing rare events, and the other class (i.e., the class with $Y=0$) as the major class, representing non-rare events.
To theoretically model the phenomenon of rare events, we adopt the idea proposed by \cite{wang2020logistic} and assume that the response probability for the minor class shrinks towards 0 as the sample size diverges to infinity.
\textcolor{black}{
Note that the rare event studied here is not an extreme value event.
In this work, a rare event refers to a random event generated by a binary distribution, where the response probability for one particular class is extremely small.
Note that the binary distribution is a discrete distribution.
In contrast, an extreme value event is typically referred to a random event generated by one particular type of extreme value distribution (e.g., Pareto distribution), which is often a continuous distribution \citep{kotz2000extreme}.
}

Rare events data are commonly encountered in practical scenarios.
For example, consider an online banking system that generates large volumes of daily transaction records.
Each transaction record can be treated as a sample and classified into two classes based on whether it is related to fraud or not.
Naturally, the probability of a transaction record being a fraud is extremely low, making fraud records rare events \citep{KRIVKO20106070,6849462}.
Another example involves computed tomography (CT) scans in medical imaging studies \citep{gu2019automatic,polat2022modified}.
The objective here is to identify specific disease regions within high-resolution CT images.
A common approach is to treat each pixel as a sample \citep{heimann2009comparison,mansoor2015segmentation}.
Subsequently, positive samples represent pixels associated with disease, while negative samples represent pixels unrelated to disease.
Often, the proportion of disease-related pixels (i.e., positive samples) is extremely small, thus  classifying them as rare events.
Other examples of rare events related data include drug discovery \citep{amaro2018ensemble,korkmaz2020deep}, data leakage prevention \citep{sigholm2012best}, and
intrusion detection in the cybersecurity scenario \citep{qian2008semi,gulati2013gmm,bagui2021resampling}.
For a comprehensive summary, we refer to \cite{chandola2009anomaly} and \cite{pimentel2014review}.

The statistical analysis of rare events data differs significantly from that of regular data.
First, rare events are rare and thus contain more valuable information than non-rare events.
For example, in a standard logistic regression model,
\cite{wang2020logistic} discovered that the statistical efficiency of the maximum likelihood estimator (MLE) is predominantly influenced by the sample size of rare events. In contrast, the sample size of non-rare events plays a considerably less significant role.
Consequently, direct implementation of the standard maximum likelihood estimation results in unnecessarily high computational costs and additional storage requirements, particularly for datasets with massive sizes.
In this regard, various undersampling methods have been developed to reduce unnecessary computation \citep{nguyen2012comparative,wang2020logistic,wang2021nonuniform}.
These methods are not limited to logistic regression and have been applied to decision trees \citep{liu2009decision,pozo2021prediction}, support vector machine \citep{4695979,bao2016boosted} and many others \citep{spelmen2018review,mohammed2020machine}.
Distributed computation methods with similar objectives have also been developed by \cite{triguero2015evolutionary,triguero2016evolutionary} and \cite{duan2020self}.

Despite significant progress in the literature on rare events data analysis, existing methods often share a common limitation: the requirement for fully labeled datasets.
This implies that the binary response $Y$ for each sample must be accurately observed and cannot be latent or missing. Consequently, a statistical learning model can be constructed to relate the binary response $Y$ to the feature vector $X$.
However, this is not always feasible in practical applications.
In fact, for many real-world applications, the collection of the feature vector $X$ is automated using well-designed hardware and software, resulting in relatively low data collection costs.
On the other hand, obtaining the response $Y$ often relies on human effort and is considerably more expensive.
For instance, in the case of obtaining a nodule mask for a lung CT image, it usually involves hiring two or more experienced radiologists to manually annotate the nodule location in a two-stage process \citep{setio2017validation}.
This makes the problem of unsupervised and semi-supervised learning a problem of great importance.

As our first attempt, we investigate one of the most commonly used unsupervised learning methods, the Gaussian Mixture Model (GMM).
The GMM is a model of fundamental importance that has been extensively studied in the literature \citep{boldea2009maximum,mclachlan2019finite}.
The Expectation-Maximization (EM) algorithm is commonly employed for GMM estimation \citep{wu1983convergence,xu1996convergence}, and its theoretical properties have been extensively examined.
Specifically, \cite{dempster1977maximum} first introduced the general form of the EM algorithm and demonstrated that the likelihood value does not decrease with each iteration.
\cite{wu1983convergence} rigorously proved that the estimator obtained by the EM algorithm numerically converges to the MLE under appropriate regularity conditions.
\cite{xu1996convergence} investigated the numerical convergence rate of the EM algorithm for GMM and established an interesting relationship between the EM algorithm and gradient ascent methods. They showed that the conditional number of the EM algorithm is always smaller than that of the gradient ascent algorithm, suggesting that the numerical convergence speed of the EM algorithm is not worse.
\cite{xu2016global} and \cite{daskalakis2017ten} conducted a global analysis of the EM algorithm for the mixture of two Gaussians from random initialization.
However, all these theoretical results were obtained under the assumption of non-rare events data.
In contrast, it has been widely observed that the standard EM algorithm exhibits painfully slow convergence rates for GMM with rare events.
This intriguing phenomenon remains unexplained by existing theories.
\cite{naim2012convergence} have provided simulation-based arguments, but no rigorously asymptotic theory has been developed in this regard.
The aim of this study is to fill this important theoretical gap.
%Motivated by this important theoretical gap, we aim to fill it with our research.

To address this problem, we formulate the EM algorithm for the GMM with rare events as an iterative algorithm governed by a contraction operator.
The convergence rate of this algorithm is predominantly determined by the spectral radius of the contraction operator.
{\color{black} Our theoretical studies reveal that, as the percentage of the rare events approaches 0, the spectral radius of the contraction operator could be arbitrarily close to 1 asymptotically,
causing the numerical convergence of the EM algorithm extremely slow.}
Our simulation studies also confirm that the spectral radius of the contraction operator approaches 1 as the response probability tends to 0.
This indicates that the numerical convergence rate of the standard EM algorithm can be extremely slow for a GMM with rare events.
To overcome this challenge, it becomes necessary to obtain a subsample of data with accurately observed labels.
By doing so, we aim to significantly enhance the numerical convergence properties of the EM algorithm. Consequently, we propose a Mixed EM (MEM) algorithm and meticulously examine its numerical convergence properties. Our findings demonstrate that the numerical convergence properties of the EM algorithm can be improved when the percentage of labeled data is carefully selected.
Extensive simulation studies are presented in this paper to demonstrate the numerical convergence properties of these two algorithms.

In summary, we aim to make the following contributions to the existing literature.
First, we theoretically prove that the numerical convergence rate of the standard EM algorithm can be extremely slow for a GMM with rare events.
Second, to fix the problem, an MEM algorithm is developed by using a partially labeled dataset.
We theoretically prove that the numerical convergence rate of this algorithm can be much faster than that of the standard EM algorithm.
These theoretical findings are further verified by extensive numerical studies.
The technical details are given in our Appendix A.1--A.2, which is heavily involved but fairly standard.
The remainder of this paper is organized as follows.
Section 2 introduces the model setting for unsupervised learning and the numerical convergence analysis of the standard EM algorithm.
Section 3 presents the model settings of semi-supervised learning and a numerical convergence analysis of the MEM algorithm.
Numerical studies are given in Section 4.
Moreover, the application of the proposed methods is then illustrated using the Swedish Traffic Signs dataset.
Finally, the article concludes with a brief discussion in Section 5.
All technical details are provided in the Appendices.

\section{Unsupervised Learning}

\subsection{Problem Setup}

Assume a total of $N$ observations are indexed by $1\le i \le N$.
The $i$th observation is denoted as $X_i\in \mR^{p}$, which is a $p$-dimensional random vector.
To model the random behavior of $X_i$, a GMM is considered \citep{mclachlan1988mixture,boldea2009maximum,mcnicholas2016mixture}.
Specifically, we assume a latent binary random variable $Y_i\in\{0,1\}$ with $P(Y_i=1)=\alpha$ for some response probability $0 < \alpha < 1$.
Next, we assume that $X_i$ follows a normal distribution with mean $\mu_0 = (\mu_{0,j}) \in \mR^{p}$ and covariance $\Sigma_0 = (\sigma_{0,j_1j_2}) \in \mR^{p\times p}$ if $Y_{i}=0$.
Otherwise, we assume another normal distribution for $X_i$ with mean $\mu_1 = (\mu_{1,j}) \in \mR^{p}$ and covariance $\Sigma_1 = (\sigma_{1,j_1j_2}) \in \mR^{p\times p}$.
Suppose $A = (A_{ij}) \in \mR^{p\times p}$ be an arbitrary square matrix of  $p\times p$ dimensions.
Then, we define $\vec(A) = (A_{ij}:1\le i, j \le p) \in \mR^{p^2}$.
Moreover, if $A$ is a symmetric matrix in the sense that $A_{ij}=A_{ji}$ for $1\le i \le j \le p$, we define the half-vec operator $\vech(A) = (A_{ij}:1\le i\le j \le p) \in \mR^{p(p+1)/2}$ \citep{boldea2009maximum}.
In this case, there should exist a unique matrix $D \in \mR^{p^2 \times p(p + 1)/2}$ such that $D\vech(A) = \vec(A)$.
We then refer to $D$ as a duplication matrix.
See \cite{magnus1988linear}, \cite{boldea2009maximum}, and \cite{magnus2019matrix} for further details.
We write $\theta = ( \alpha, \mu_0^\top, \vech(\Sigma_0)^\top, \mu_1^\top, \vech(\Sigma_1)^\top )^\top \in \mR^{q}$ as our interested parameter vector with $q = p^2+3p+1$.
Then, the probability density function of the GMM can be written as
$f_\theta(x) = \alpha \phi_{\mu_1,\Sigma_1}(x) + (1-\alpha) \phi_{\mu_0,\Sigma_0}(x)$,
where $\phi_{\mu,\Sigma}(x) = (|2\pi\Sigma|)^{-1/2}\exp\{-(x-\mu)^\top \Sigma^{-1} (x-\mu)/2 \}$ is the probability density function of a multivariate normal distribution with mean $\mu \in \mR^p$ and covariance $\Sigma \in \mR^{p\times p}$ \citep{mclachlan1988mixture}.

For a classical GMM, we should have $\alpha \in (0,1)$ as a fixed parameter in the sense that it does not vanish as the sample size increases.
Under this setup, we should have the expected sample size ratio $\color{black} E(N_1)/N =\alpha \nrightarrow 0$ as $N \to \infty$, where $N_1 = \sum_{i=1}^N Y_i$.
Unfortunately, this classical setup is inappropriate for applications involving rare events.
In this case, one of the two latent classes (e.g., the class with $Y_i=1$) is considered as the minor class with a mixing probability $\alpha \to 0$ as $N \to \infty$, and the other latent class (e.g., the class with $Y_i = 0$) is considered as the major class with a mixing probability $1-\alpha \to 1$ as $N \to \infty$.
Consequently, the sample size associated with the minor class (e.g., $N_1$) is significantly smaller than that of the major class (e.g., $N_0=N-N_1$) in the sense that $N_1/N \to_p 0$ as $N \to \infty$.
It is worthwhile mentioning that $\alpha$ cannot be excessively small either.
Otherwise, the sample size of the minor latent class (i.e., $N_1$) might shrink towards 0.
Consequently, the parameters associated with the minor class (i.e., $\mu_1$ and $\Sigma_1$) cannot be consistently estimated even if class label $Y_i$s are given.
In this case, no meaningful asymptotic theory can be established.
For the sake of asymptotic theory development, we do need to assume that $E(N_1)\to\infty$ as $N\to\infty$.
This further requires that $N\alpha\to\infty$ as $N\to\infty$.
Therefore, throughout the remainder of this article, we always assume that (1) $\alpha\to 0$ and (2) $N\alpha \to \infty$ as $N \to \infty$; see \cite{wang2020logistic}, \cite{wang2021nonuniform} and \cite{li2023distributed} for further discussion.
These two assumptions are called the rare events assumptions.

% \newpage

\subsection{The EM Algorithm}

Next, we consider how to compute the MLE for the GMM.
The log-likelihood function is given as follows
\beqr
\label{eq: GMM likelihood function}
\mL(\theta) = \sum_{i=1}^N \log f_\theta(X_i).
\eeqr
Subsequently, we can obtain the MLE as $\wh\theta = ( \wh\alpha, \wh\mu_0^\top, \vech(\wh\Sigma_0)^\top, \wh\mu_1^\top, \vech(\wh\Sigma_1)^\top )^\top = \argmax_{\theta}$
$\mL (\theta)$.
To compute the MLE, a classical EM algorithm can be used \citep{dempster1977maximum,biernacki2003choosing}.
In the past literature, a standard EM algorithm was often motivated by the method of complete log-likelihood, where the latent class label $Y_i$s are pretended to be known.
In this study, we present another interesting perspective, where the EM algorithm is inspired by the gradient condition of the log-likelihood function without knowing the latent class labels.
%By doing so, we can establish a more clear relationship between the EM algorithm and the gradient directions.

{\color{black}
Specifically, define $\dot\mL(\theta) = ( \dot\mL_\alpha(\theta), \dot\mL_{\mu_0}(\theta)^\top, \dot\mL_{\Sigma_0}(\theta)^\top, \dot\mL_{\mu_1}(\theta)^\top, \dot\mL_{\Sigma_1}(\theta)^\top )^\top \in \mR^{q}$, where $\dot\mL_\alpha(\theta)=\partial\mL(\theta)/\partial\alpha \in \mR$, $\dot\mL_{\mu_k}(\theta)=\partial\mL(\theta)/\partial\mu_k \in \mR^p$ and
$\dot\mL_{\Sigma_k}(\theta) = \partial\mL(\theta)/ \partial \vech(\Sigma_k)\in \mR^{p(p+1)/2}$ with $k \in \{0,1\}$.
Recall that $\wh\theta$ denotes the MLE.
Consequently, we should have a gradient condition as $\dot\mL(\wh\theta)=0$.
This suggests a set of estimation equations as
$\wh\alpha = \mF_\alpha(\wh\theta)$,
$\wh\mu_k = \mF_{\mu_k}(\wh\theta)$ and
$\vech(\wh\Sigma_k) = \mF_{\Sigma_k}(\wh\theta)$ with $k \in \{0,1\}$.
Here $\mF(\theta) =$
$(\mF_{\alpha}(\theta), \mF_{\mu_0}(\theta)^\top, \mF_{\Sigma_0}(\theta)^\top$, $\mF_{\mu_1}(\theta)^\top, \mF_{\Sigma_1} (\theta)^\top )^\top \in \mR^q$ is a mapping function, where
\begin{gather}
\mF_\alpha(\theta) = N^{-1}\sum_{i=1}^N \pi_i \in \mR,
\ \mF_{\mu_0}(\theta) = \Big\{ \sum_{i=1}^N (1-\pi_i) \Big\}^{-1}  \sum_{i=1}^N (1-\pi_i) X_i \in \mR^{p}, \nonumber \\
\mF_{\Sigma_0}(\theta) = \Big\{ \sum_{i=1}^N (1-\pi_i) \Big\}^{-1}  \sum_{i=1}^N \Big(1-\pi_i\Big) \vech\Big\{ \Big( X_i-\mu_0 \Big) \Big( X_i-\mu_0 \Big)^\top \Big\} \in \mR^{\frac{(p+1)p}{2}},
\nonumber \\
\mF_{\mu_1}(\theta) = \Big( \sum_{i=1}^N \pi_i \Big)^{-1} \sum_{i=1}^N \pi_i X_i \in \mR^{p}, \label{eq: mapping function} \\
\mF_{\Sigma_1}(\theta) = \Big( \sum_{i=1}^N \pi_i \Big)^{-1} \sum_{i=1}^N \pi_i \vech\Big\{ \Big( X_i-\mu_1 \Big) \Big( X_i-\mu_1 \Big)^\top \Big\} \in \mR^{\frac{(p+1)p}{2}}, \nonumber
\end{gather}
and $\pi_i =P(Y_i=1 | X_i) = f_\theta(X_i)^{-1} \alpha \phi_{\mu_1,\Sigma_1}(X_i)$ is the posterior probability.
An iterative algorithm can then be developed accordingly.
% This leads to an iterative algorithm as follows.

Let $\wh\theta^{(0)} = ( \wh\alpha^{(0)}, \wh\mu_0^{(0) \top}, \vech(\wh\Sigma_0^{(0)})^\top, \wh\mu_1^{(0) \top}$, $\vech(\wh\Sigma_1^{(0)})^\top )^\top  \in \mR^{q}$ be the initial estimator.
Let $\wh\theta^{(t)} = ( \wh\alpha^{(t)}, \wh\mu_0^{(t) \top}, \vech(\wh\Sigma_0^{(t)})^\top, \wh\mu_1^{(t) \top}$, $\vech(\wh\Sigma_1^{(t)})^\top )^\top $ $\in \mR^{q}$ be the estimator obtained in the $t$th step.
Subsequently, we obtain the $(t+1)$th step estimator $\wh\theta^{(t+1)}= ( \wh\alpha^{(t+1)}, \wh\mu_0^{(t+1) \top}, \vech(\wh\Sigma_0^{(t+1)})^\top$,
$\wh\mu_1^{(t+1) \top}, \vech(\wh\Sigma_1^{(t+1)})^\top )^\top  \in \mR^{q}$ as
$\wh\alpha^{(t+1)} = \mF_\alpha(\wh\theta^{(t)})$, $\wh\mu_k^{(t+1)} = \mF_{\mu_k}(\wh\theta^{(t)})$ and $\vech(\wh\Sigma_k^{(t+1)}) = \mF_{\Sigma_k}(\wh\theta^{(t)})$ with $k \in \{0,1\}$.}
% and $\wh\pi_i^{(t+1)} = f_{\wh\theta^{(t)}}(X_i)^{-1} \wh\alpha^{(t)} \phi_{\wh\mu_1^{(t)},\wh\Sigma_1^{(t)}}(X_i)$.
The aforementioned steps should be iteratively executed until convergence.
By the time of convergence, we obtain the MLE as $\wh\theta = ( \wh\alpha, \wh\mu_0^\top, \vech(\wh\Sigma_0)^\top, \wh\mu_1^\top$, $\vech(\wh\Sigma_1)^\top )^\top \in \mR^q$.
As one can see, this is an algorithm fully implied by the gradient condition.
{\color{black} Another possible way to develop an algorithm is the complete-data likelihood method, where the latent class labels $Y_i$s are assumed to be known\citep{dempster1977maximum,wu1983convergence}.
This leads to a set of maximum likelihood estimators for the interested parameters with analytical formula.
This constitutes the maximization step.
Once a set of estimators are obtained for the interested parameters, the posterior probabilities of the latent class label can be analytically derived.
This constitutes the expectation step.
Both the maximization and expectation steps lead to a standard EM algorithm, as to be demonstrated in the first part of Appendix A.3.
We find that this standard EM algorithm is exactly the same as the above algorithm inspired by the gradient condition \citep{dempster1977maximum,wu1983convergence}.}

\subsection{Numerical Convergence Analysis}

To investigate the numerical convergence property, the key issue is to study the difference between $\wh\theta$ and $\wh\theta^{(t+1)}$.
Subsequently, the EM algorithm can be simply written as $\wh\theta^{(t+1)} = \mF(\wh\theta^{(t)})$ by \eqref{eq: mapping function}.
If $\wh\theta^{(t)}$ converges numerically to $\wh\theta$ as $t\to\infty$, we should have $\wh\theta = \mF(\wh\theta)$.
Consequently, we have {\color{black} $\wh\theta^{(t+1)} - \wh\theta = \mF(\wh\theta^{(t)})-\mF(\wh\theta)$.
Define $\mF_j(\theta)$ as the $j$th element of the mapping function $\mF(\theta)$.
We then have $\mF_j(\wh\theta^{(t)})-\mF_j(\wh\theta) = \dot\mF_j(\wt\theta^{(t)}_j) (\wh\theta^{(t)}-\wh\theta)$, where $\wt\theta^{(t)}_j= \eta^{(t)}_j \wh\theta^{(t)} + (1-\eta^{(t)}_j)\wh\theta \in \mR^q$ and $0<\eta^{(t)}_j<1$ \citep{feng2013mean}.}
We define $\dot\mF(\theta) = (\dot\mF_{\alpha}(\theta), \dot\mF_{\mu_0}(\theta)^\top,
\dot\mF_{\Sigma_0}(\theta)^\top, \dot\mF_{\mu_1}(\theta)^\top, \dot\mF_{\Sigma_1}(\theta)^\top )^\top \in \mR^{q\times q}$ as the contraction operator,
where $\dot\mF_{\alpha}(\theta)= \partial\mF_{\alpha}(\theta)/\partial \theta =( \dot{\mF}_{\alpha\alpha} (\theta), \dot{\mF}_{\alpha\mu_0} (\theta)^\top,
\dot{\mF}_{\alpha\Sigma_0} (\theta)^\top,$ $\dot{\mF}_{\alpha\mu_1} (\theta)^\top, \dot{\mF}_{\alpha\Sigma_1} (\theta)^\top )^\top \in \mR^q$, $\dot\mF_{\mu_k}(\theta)
= \partial\mF_{\mu_k}(\theta)/\partial \theta^\top =(\dot{\mF}_{\mu_k\alpha} (\theta),
\dot{\mF}_{\mu_k\mu_0} (\theta), \dot{\mF}_{\mu_k\Sigma_0} (\theta), \dot{\mF}_{\mu_k\mu_1} (\theta)$, $\dot{\mF}_{\mu_k\Sigma_1} (\theta) ) \in \mR^{p\times q}$, and $\dot\mF_{\Sigma_k}(\theta) = \partial\mF_{\Sigma_k}(\theta)/\partial \theta^\top
= (\dot{\mF}_{\Sigma_k\alpha} (\theta),
\dot{\mF}_{\Sigma_k\mu_0} (\theta), \dot{\mF}_{\Sigma_k\Sigma_0} (\theta), \dot{\mF}_{\Sigma_k\mu_1} (\theta),$ $\dot{\mF}_{\Sigma_k\Sigma_1} (\theta) ) \in \mR^{p(p+1)/2\times q}$ with $k \in \{0,1\}$.
Therefore, the asymptotic behavior of $\dot\mF(\theta)$
% $\color{black} \dot\mF(\wt\theta^{(t)}) \approx \dot\mF(\theta)$
is critical for the numerical convergence of the EM algorithm \citep{xu1996convergence,6790184}.
Next, we should study the asymptotic behavior of $\dot\mF(\theta)$ under the rare events assumptions with great care.
This leads to the following theorem. Its rigorous proof is provided in Appendix A.1.

\bet
\label{Theorem 1}
Assume $\alpha \to 0$ and $N\alpha\to \infty$ as $N\to\infty$. We then have as $N\to\infty$
\beqrs
\dot\mF(\theta) \to_p \mM =
\left(\begin{array}{ccccc}
1 & 0 & 0 & 0 & 0 \\
\mu_0-\mu_1 & 0 & 0 & 0 & 0 \\
\Delta_{\Sigma_0\alpha} & 0 & 0 & 0 & 0 \\
\Delta_{\mu_1\alpha} & - \Sigma_1\Sigma_0^{-1} & \Delta_{\mu_1\Sigma_0} & I_p & \Delta_{\mu_1\Sigma_1} \\
\vech(\Delta_{\Sigma_1\alpha}) & \Delta_{\Sigma_1\mu_0} & \Delta_{\Sigma_1\Sigma_0} & \Delta_{\Sigma_1\mu_1} & -\Delta_{\Sigma_1\Sigma_1} \\
\end{array}\right),
\eeqrs
where $\Delta_{\Sigma_0\alpha} = -\vech\{ \Sigma_1-\Sigma_0+(\mu_1-\mu_0)(\mu_1-\mu_0)^\top \} \in \mR^{p(p+1)/2}$,
$\Delta_{\mu_1\alpha} = (\Delta_{\mu_1\alpha,j}) \in \mR^{p}$, $\Delta_{\mu_1\alpha,j} = - \int \phi_{\mu_1,\Sigma_1}^2(x) \phi_{\mu_0,\Sigma_0}^{-1}(x) (x_j-\mu_{1,j}) dx$,
$\Delta_{\mu_1\Sigma_k} = (-1)^{k+1} E\{ ( X_i - \mu_1 ) \vec (\gamma_{ik}\gamma_{ik}^\top )^\top |Y_i=1 \} D/2 \in \mR^{p \times p(p+1)/2}$ with $k \in \{0,1\}$,
$\Delta_{\Sigma_1\alpha} = (\Delta_{\Sigma_1\alpha,j_1j_2})\in \mR^{p\times p}$,
$\Delta_{\Sigma_1\alpha,j_1j_2} = - \int \phi_{\mu_1,\Sigma_1}^2(x)$ $\phi_{\mu_0,\Sigma_0}^{-1} (x) \{ (x_{j_1}-\mu_{1,j_1}) (x_{j_2}-\mu_{1,j_2})-\sigma_{1,j_1j_2} \} dx$,
$\Delta_{\Sigma_1\mu_k} = (-1)^{k+1} E [ \vech \{ (X_i-\mu_1)(X_i-\mu_1)^\top \} \gamma_{ik}^\top | Y_i=1] - (-1)^{k+1} \vech (\Sigma_1) (\mu_1-\mu_k)^\top \Sigma_k^{-1} \in \mR^{p(p+1)/2 \times p}$ with $k \in \{0,1\}$,
$\Delta_{\Sigma_1\Sigma_k}
= E [ \vech \{ ( X_i-\mu_1 )( X_i-\mu_1 )^\top \} \vec( \gamma_{ik}\gamma_{ik}^\top )^\top | Y_i=1] D/2
- \vech (\Sigma_1 ) \vec[ \Sigma_k^{-1}
\{ \Sigma_1+ (\mu_1-\mu_k)(\mu_1-\mu_k)^\top \} \Sigma_k^{-1} ]^\top D/2 \in \mR^{\frac{p(p+1)}{2} \times \frac{p(p+1)}{2}}$ and $\gamma_{ik} = \Sigma_k^{-1} (X_i-\mu_k)$ with $k \in \{0,1\}$.
\eet
\noindent
By Theorem \ref{Theorem 1}, we obtain a number of interesting findings.
{\color{black}First, it can be confirmed that the determinant of $\mM - I_q$ is 0.
Therefore, $\mM$ has at least one eigenvalue as 1.
This further implies that the spectral radius of the contraction operator is no less than 1 as $\alpha \to 0$ asymptotically.
This suggests that the numerical convergence rate of the EM algorithm for GMM with rare events could be extremely slow, thereby necessitating the large number of iterations.
More specific, we} know that $\wh\alpha^{(t+1)}-\wh\alpha \approx \wh\alpha^{(t)}-\wh\alpha$, since the first diagonal component of $\mM$ is 1 whereas all other components in the first row are 0.
This suggests that the numerical convergence rate of $\wh\alpha^{(t)}$ should be extremely slow.
Unfortunately, this side effects spill over into both $\wh\mu_0^{(t)}$ and $\wh\Sigma_0^{(t)}$.
Specifically, by Theorem 1, we know that $\wh\mu_0^{(t+1)}-\wh\mu_0 \approx (\mu_0-\mu_1) (\wh\alpha^{(t)}-\wh\alpha)$, since the $(2,1)$th component of $\mM$ is $(\mu_0-\mu_1)$.
This suggests that the numerical convergence rate of $\wh\mu_0^{(t)}$ is mainly controlled by $\wh\alpha^{(t)}$, which unfortunately converges at an extremely slow speed as mentioned before.
Therefore, we know that the numerical convergence rate of $\wh\mu_0^{(t)}$ cannot be fast either.
By similar arguments, we know that the numerical convergence property of $\wh\Sigma_0^{(t)}$ is not optimistic either.
The numerical convergence properties of $\wh\mu_1^{(t)}$ and $\wh\Sigma_1^{(t)}$ are considerably more complicated.
To gain an intuitive understanding, we consider an ideal case, where $\wh\alpha^{(t)} = \wh\alpha$, $\wh\mu_0^{(t)} = \wh\mu_0$, and $\wh\Sigma_k^{(t)} = \wh\Sigma_k$ with $k \in \{ 0,1\}$.
Then by the 4th row of $\mM$, we know that $\wh\mu_1^{(t+1)}- \wh\mu_1 \approx \wh\mu_1^{(t)} - \wh\mu_1$, which implies that $\wh\mu_1^{(t)}$ cannot converge fast, even if the maximum likelihood estimators $\wh\alpha$, $\wh\mu_0$ and $\wh\Sigma_k$ are already given.
This side effect also spills over into the numerical convergence property of $\wh\Sigma_1$, since the numerical convergence of $\wh\Sigma_1^{(t)}$ is also affected by that of $\wh\mu_1^{(t)}$.
The consequence is that $\wh\theta^{(t+1)}$ ceases approaching $\wh\theta$.
Then how to fix the problem becomes an important problem.

\section{Semi-Supervised Learning}

\subsection{Partially Labeled Sample}

By the careful analysis of the previous section, we know that the MLE of a GMM with rare events can hardly be computed by a standard EM algorithm, if only unlabeled data are available.
To fix the problem, it seems very necessary to obtain an additional subsample of data with accurately observed labels.
By doing so, we wish the numerical convergence properties of the EM algorithm can be improved substantially.
{\color{black} Note that the sole purpose for comparing the MEM and EM algorithms is not
to establish any type of superiority about the MEM algorithm over the EM algorithm.
This comparison is obviously unfair since the MEM algorithm enjoys the information provided by partially labeled instances while the EM algorithm does not.
The sole purpose here is to demonstrate the value of those partially labeled instances.
We are able to do so since the only difference between the MEM and EM algorithms is the partially labeled data.
Therefore, any performance improvements as demonstrated by the MEM algorithm over the EM algorithm can be fully attributed to those partially labeled instances.}

Specifically, we assume that the size of this labeled dataset is $m$ with $m/N \to \kappa$ as $N \to \infty$ for some $\kappa>0$.
Next, for each labeled sample $i$, we use $Y_i^* \in \{0,1\}$ to represent the observed response with $P(Y_i^*=1) = \alpha$.
{\color{black}
Here we assume that the distribution of the labeled data is the same as that of the unlabeled data.}
Lastly, we use $X_i^*$ to represent the associated feature vector.
Here the conditional distribution of $X_i^*$ given $Y_i^*$ should remain the same as that of the unlabeled data.
Subsequently, a joint log-likelihood function can be analytically spelled out as
\begin{gather}
\mL_{\rm semi} (\theta) = \sum_{i=1}^N \log f_\theta \big(X_i\big)
+ \sum_{i=1}^m \bigg\{ Y_i^* \log \phi_{\mu_1,\Sigma_1} \big(X_i^*\big) + \Big(1-Y_i^*\Big) \log \phi_{\mu_0,\Sigma_0} \big(X_i^*\big) \bigg\} \nonumber \\
+ \sum_{i=1}^m \bigg\{ Y_i^* \log \alpha + \Big(1-Y_i^*\Big) \log \big(1-\alpha\big) \bigg\}.
\label{eq: mixed likelihood function}
\end{gather}
Comparing this log-likelihood function with \eqref{eq: GMM likelihood function}, the key difference is that the labeled samples are partially involved in \eqref{eq: mixed likelihood function} but not in \eqref{eq: GMM likelihood function}.
Accordingly, a new estimator can be defined as $\wh\theta_{\rm semi} = ( \wh\alpha^{\rm semi}, \wh\mu_0^{\rm semi \top}, \vech(\wh\Sigma_{0}^{\rm semi} )^\top, \wh\mu_1^{\rm semi \top}, \vech(\wh\Sigma_1^{\rm semi})^\top )^\top = \argmax_{\theta} \mL_{\rm semi} (\theta)$.

% \newpage

\subsection{The MEM Algorithm}

{\color{black}
Next, we consider how to compute the MLE for the new log-likelihood function \eqref{eq: mixed likelihood function}.
Specifically, define $\dot\mL_{\rm semi}(\theta) = ( \dot\mL_{{\rm semi}, \alpha}(\theta), \dot\mL_{{\rm semi}, \mu_0}(\theta)^\top$, $\dot\mL_{{\rm semi}, \Sigma_0}(\theta)^\top$, $\dot\mL_{{\rm semi}, \mu_1}(\theta)^\top$, $\dot\mL_{{\rm semi}, \Sigma_1}(\theta)^\top )^\top$ $\in \mR^{q}$, where $\dot\mL_{{\rm semi}, \alpha}(\theta) = \partial \mL_{\rm semi}(\theta)/\partial\alpha \in \mR$, $\dot\mL_{{\rm semi}, \mu_k}(\theta) = \partial \mL_{\rm semi}(\theta)/\partial\mu_k \in \mR^p$ and $\dot\mL_{{\rm semi}, \Sigma_k}(\theta)$ $= \partial\mL_{\rm semi}(\theta)/ \partial \vech (\Sigma_k) \in \mR^{p(p+1)/2}$ with $k \in \{0,1\}$.
Recall that $\wh\theta_{\rm semi}$ is the MLE.
We should have a gradient condition as $\dot\mL_{\rm semi}(\wh\theta_{\rm semi})=0$.
This suggests a set of estimation equations as
$\wh\alpha^{\rm semi} = \mF^*_\alpha(\wh\theta_{\rm semi})$, $\wh\mu_k^{\rm semi} = \mF^*_{\mu_k}(\wh\theta_{\rm semi})$ and $\vech(\wh\Sigma_k^{\rm semi}) = \mF^*_{\Sigma_k}(\wh\theta_{\rm semi})$ with $k \in \{0,1\}$.
% where $\wh\pi_i^{\rm semi} = f_{\wh\theta_{\rm semi}}(X_i)^{-1} \wh\alpha \phi_{\wh\mu_1^{\rm semi},\wh\Sigma_1^{\rm semi}}(X_i)$.
Here $\mF^*(\theta) = ( \mF^*_\alpha(\theta), \mF^*_{\mu_0}(\theta)^\top$, $\mF^*_{\Sigma_0}(\theta)^\top, \mF^*_{\mu_1}(\theta)^\top, \mF^*_{\Sigma_1}(\theta)^\top )^\top$
$\in \mR^q$ is defined as a mapping function, where
\begin{gather}
\mF^*_\alpha(\theta) = \Big(N+m\Big)^{-1} \Big( \sum_{i=1}^N \pi_i + \sum_{i=1}^m Y_i^* \Big) \in \mR, \nonumber \\
\mF^*_{\mu_0}(\theta) = \Big\{ \sum_{i=1}^N \Big( 1-\pi_i \Big) + \sum_{i=1}^m \Big(1-Y_i^*\Big) \Big\}^{-1}  \Big\{ \sum_{i=1}^N \Big(1-\pi_i \Big) X_i
+ \sum_{i=1}^m \Big(1-Y_i^*\Big) X_i^* \Big\} \in \mR^{p}, \nonumber \\
\mF^*_{\Sigma_0}(\theta) = \Big\{ \sum_{i=1}^N \Big(1-\pi_i\Big)+\sum_{i=1}^m \Big(1-Y_i^*\Big) \Big\}^{-1} \Big[ \sum_{i=1}^N \Big(1-\pi_i \Big)
\vech\Big\{ \Big( X_i-\mu_0 \Big) \Big( X_i -\mu_0 \Big)^\top \Big\} \nonumber \\
+ \sum_{i=1}^m \Big(1-Y_i^*\Big) \vech\Big\{ \Big( X_i^*-\mu_0 \Big) \Big( X_i^*-\mu_0 \Big)^\top \Big\} \Big] \in \mR^{\frac{p(p+1)}{2}},
\label{eq: mapping function semi} \\
\mF^*_{\mu_1}(\theta)
= \Big( \sum_{i=1}^N \pi_i  +\sum_{i=1}^m Y_i^* \Big)^{-1} \Big( \sum_{i=1}^N \pi_i X_i + \sum_{i=1}^m Y_i^* X_i^* \Big) \in \mR^{p}, \nonumber
\end{gather}
\begin{gather}
\mF^*_{\Sigma_1}(\theta) = \Big( \sum_{i=1}^N \pi_i +\sum_{i=1}^m Y_i^* \Big)^{-1} \Big[ \sum_{i=1}^N \pi_i \vech\Big\{ \Big( X_i-\mu_1 \Big) \Big( X_i-\mu_1 \Big)^\top \Big\} \nonumber \\
+ \sum_{i=1}^m Y_i^* \vech\Big\{ \Big( X_i^*-\mu_1 \Big) \Big( X_i^*-\mu_1 \Big)^\top \Big\} \Big] \in \mR^{\frac{p(p+1)}{2}}, \nonumber
\end{gather}
and $\pi_i =P(Y_i=1|X_i)= f_\theta(X_i)^{-1} \alpha \phi_{\mu_1,\Sigma_1}(X_i)$ is the posterior probability.
This leads to an iterative algorithm as follows.

Let $\wh\theta_{\rm semi}^{(0)} = ( \wh\alpha^{{\rm semi} (0)}, \wh\mu_0^{{\rm semi} (0) \top}, \vech(\wh\Sigma_0^{{\rm semi} (0)})^\top, \wh\mu_1^{{\rm semi} (0) \top}$, $\vech(\wh\Sigma_1^{{\rm semi} (0)})^\top )^\top \in \mR^{q}$ be the initial estimator.
Let $\wh\theta_{\rm semi}^{(t)} = ( \wh\alpha^{{\rm semi} (t)}, \wh\mu_0^{{\rm semi} (t) \top}, \vech(\wh\Sigma_0^{{\rm semi} (t)})^\top$, $\wh\mu_1^{{\rm semi} (t) \top}, \vech(\wh\Sigma_1^{{\rm semi} (t)})^\top)^\top$ $\in \mR^{q}$ be the estimator obtained in the $t$th step.
Consequently, the next step estimator $\wh\theta_{\rm semi}^{(t+1)}= ( \wh\alpha^{{\rm semi} (t+1)}, \wh\mu_0^{{\rm semi} (t+1) \top}$, $\vech(\wh\Sigma_0^{{\rm semi} (t+1)})^\top, \wh\mu_1^{{\rm semi} (t+1) \top}$, $\vech$ $(\wh\Sigma_1^{{\rm semi} (t+1)})^\top )^\top \in \mR^{q}$ can be obtained as
$\wh\alpha^{{\rm semi} (t+1)} = \mF^*_\alpha(\wh\theta_{\rm semi}^{(t)})$,
$\wh\mu_k^{{\rm semi} (t+1)} = \mF^*_{\mu_k}(\wh\theta_{\rm semi}^{(t)})$,
$\vech(\wh\Sigma_k^{{\rm semi} (t+1)}) = \mF^*_{\Sigma_k}(\wh\theta_{\rm semi}^{(t)})$
and $\wh\pi_i^{{\rm semi} (t+1)} = f_{\wh\theta_{\rm semi}^{(t)}}(X_i)^{-1} \wh\alpha^{{\rm semi} (t)}$ $\phi_{\wh\mu_1^{{\rm semi} (t)},\wh\Sigma_1^{{\rm semi} (t)}}(X_i)$ with $k \in \{0,1\}$.}
The aforementioned steps should be iteratively executed until convergence.
By the time of convergence, we obtain the MLE as $\wh\theta_{\rm semi} = ( \wh\alpha^{\rm semi}, \wh\mu_0^{\rm semi \top}, \vech(\wh\Sigma_{0}^{\rm semi} )^\top$, $\wh\mu_1^{\rm semi \top}$, $\vech(\wh\Sigma_1^{\rm semi})^\top )^\top$ $\in \mR^{q}$.
{\color{black} Note that this is also an EM-type algorithm but with mixed samples, in the sense that some samples are labeled while the others are not.
For convenience, we refer to this interesting algorithm as an MEM algorithm.
As one can see, this MEM algorithm presented above is inspired by the gradient condition $\dot\mL_{\rm semi}(\wh\theta_{\rm semi})=0$.
In fact, it can also be motivated by the complete-data likelihood method, which  assumes that the latent $Y_i$s are actually observed.
This leads to a standard EM algorithm; see the second part of Appendix A.3 for a brief description, which is the same as the MEM algorithm.
Therefore, the MEM algorithm is indeed a standard EM algorithm \citep{dempster1977maximum,wu1983convergence}.}

\subsection{Numerical Convergence Analysis}

To investigate the numerical convergence property, the key issue is to study the difference between $\wh\theta_{\rm semi}$ and $\wh\theta_{\rm semi}^{(t+1)}$.
By \eqref{eq: mapping function semi}, the MEM algorithm can be written as $\wh\theta_{\rm semi}^{(t+1)} = \mF^*(\wh\theta_{\rm semi}^{(t)})$.
If $\wh\theta_{\rm semi}^{(t)}$ converges numerically to $\wh\theta_{\rm semi}$ as $t\to\infty$, we then should have $\wh\theta_{\rm semi} = \mF^*(\wh\theta_{\rm semi})$.
Consequently, we have $\wh\theta_{\rm semi}^{(t+1)} - \wh\theta_{\rm semi} = \mF^*(\wh\theta_{\rm semi}^{(t)})-\mF^*(\wh\theta_{\rm semi})$.
Define $\mF_j^*(\theta)$ as the $j$th element of the mapping function $\mF^*(\theta)$.
We then have $\mF^*_j(\wh\theta_{\rm semi}^{(t)})-\mF^*_j(\wh\theta_{\rm semi})$
$= \dot\mF^*_j(\wt\theta_j^{* (t)}) (\wh\theta_{\rm semi}^{(t)}-\wh\theta_{\rm semi})$,
where $\wt\theta_j^{* (t)}= \eta^{* (t)}_j \wh\theta_{\rm semi}^{(t)} + (1-\eta^{* (t)}_j)\wh\theta_{\rm semi} \in\mR^q$ and $0<\eta^{* (t)}_j<1$ \citep{feng2013mean}.
Here define $\dot\mF^*(\theta)$ $= (\dot\mF^*_{\alpha}(\theta), \dot\mF^*_{\mu_0}(\theta)^\top,
\dot\mF^*_{\Sigma_0}(\theta)^\top, \dot\mF^*_{\mu_1}(\theta)^\top, \dot\mF^*_{\Sigma_1}(\theta)^\top )^\top \in \mR^{q\times q}$ as the contraction operator,
where $\dot\mF^*_{\alpha}(\theta)= \partial\mF^*_{\alpha}(\theta)/\partial \theta =( \dot{\mF^*}_{\alpha\alpha} (\theta), \dot{\mF^*}_{\alpha\mu_0} (\theta)^\top, \dot{\mF^*}_{\alpha\Sigma_0} (\theta)^\top$,
$\dot{\mF^*}_{\alpha\mu_1} (\theta)^\top, \dot{\mF^*}_{\alpha\Sigma_1} (\theta )^\top )^\top$ $\in \mR^q$, $\dot\mF^*_{\mu_0}(\theta)
= \partial\mF^*_{\mu_0}(\theta)/\partial \theta^\top =(\dot{\mF^*}_{\mu_0\alpha} (\theta),
\dot{\mF^*}_{\mu_0\mu_0} (\theta), \dot{\mF^*}_{\mu_0\Sigma_0} (\theta)$, $\dot{\mF^*}_{\mu_0\mu_1} (\theta), \dot{\mF^*}_{\mu_0\Sigma_1} (\theta) ) $
$\in \mR^{p\times q}$ and $\dot\mF^*_{\Sigma_0}(\theta) = \partial\mF^*_{\Sigma_0}(\theta)/\partial \theta^\top = (\dot{\mF^*}_{\Sigma_0\alpha} (\theta),
\dot{\mF^*}_{\Sigma_0\mu_0} (\theta)$, $\dot{\mF^*}_{\Sigma_0\Sigma_0} (\theta)$, $\dot{\mF^*}_{\Sigma_0\mu_1} (\theta)$,
$\dot{\mF^*}_{\Sigma_0\Sigma_1}$
$(\theta) ) \in \mR^{\frac{p(p+1)}{2}\times q}$,
$\dot\mF^*_{\mu_1}(\theta)
%= \partial\mF^*_{\mu_1}(\theta)/\partial \theta
=( \alpha \dot{\mF^*}_{\mu_1\alpha} (\theta),
\dot{\mF^*}_{\mu_1\mu_0} (\theta), \dot{\mF^*}_{\mu_1\Sigma_0} (\theta)$,
$\dot{\mF^*}_{\mu_1\mu_1} (\theta)$, $\dot{\mF^*}_{\mu_1\Sigma_1} (\theta) ) \in \mR^{p\times q}$ and $\dot\mF^*_{\Sigma_1}(\theta) $
%$= \partial\mF^*_{\Sigma_1}(\theta)/\partial \theta $
$= ( \alpha \dot{\mF^*}_{\Sigma_1\alpha} (\theta)$,
$\dot{\mF^*}_{\Sigma_1\mu_0} (\theta)$, $\dot{\mF^*}_{\Sigma_1\Sigma_0} (\theta)$, $\dot{\mF^*}_{\Sigma_1\mu_1} (\theta)$, $\dot{\mF^*}_{\Sigma_1\Sigma_1} (\theta) )$ $\in \mR^{\frac{p(p+1)}{2}\times q}$.
Hence, the asymptotic behavior of
$\dot\mF^*(\theta)$
% $\dot\mF^*(\wt\theta_{\rm semi}^{(t)}) \approx \dot\mF^*(\theta)$
is important for the numerical convergence of the EM algorithm \citep{xu1996convergence,6790184}.
Subsequently, we should study the asymptotic behavior of $\dot\mF^*(\theta)$ under the rare events assumption with great care.
This leads to the following theorem. Its rigorous proof is provided in Appendix A.2.

\bet
\label{Theorem 2}
Assume $\alpha \to 0$ and $N\alpha\to \infty$ as $N\to\infty$.
Assume $\lim_{N\to\infty} m/N = \kappa$.
We then have $\dot\mF^*(\theta) \to_p \mM^*$ as $N\to\infty$, where
\begin{gather}
\mM^* =
\left(\begin{array}{ccccc}
(1+\kappa )^{-1} & 0 & 0 & 0 & 0 \\
\frac{\mu_0-\mu_1}{1+\kappa} & 0 & 0 & 0 & 0 \\
\frac{\Delta_{\Sigma_0\alpha}}{1+\kappa} & 0 & 0 & 0 & 0 \\
0 & \frac{- \Sigma_1\Sigma_0^{-1}}{1+\kappa} & \frac{\Delta_{\mu_1\Sigma_0}}{1+\kappa} & \frac{I_p}{1+\kappa}  & \frac{\Delta_{\mu_1\Sigma_1}}{1+\kappa} \\
0 & \frac{\Delta_{\Sigma_1\mu_0}}{1+\kappa}  & \frac{\Delta_{\Sigma_1\Sigma_0}}{1+\kappa} & \frac{\Delta_{\Sigma_1\mu_1}}{1+\kappa} & -\frac{\Delta_{\Sigma_1\Sigma_1}}{1+\kappa} \\
\end{array}\right). \nonumber
\end{gather}
\eet
%\noindent

By Theorem \ref{Theorem 2}, several interesting findings can be obtained.
First, we know that $\wh\alpha^{{\rm semi}(t+1)}-\wh\alpha^{\rm semi} \approx (1+\kappa)^{-1} (\wh\alpha^{{\rm semi} (t)}-\wh\alpha^{\rm semi})$, since the first diagonal component of $\mM^*$ is $(1+\kappa)^{-1}$ whereas all other components in the first row are 0.
This suggests that $\wh\alpha^{{\rm semi} (t)}$ converges linearly as long as $\kappa>0$.
Consequently, the slow convergence rate problem
%of $\wh\alpha^{{\rm semi} (t)}$
as described in Theorem \ref{Theorem 1} is nicely rectified.
Meanwhile, we know that $\wh\mu_0^{{\rm semi} (t+1)}-\wh\mu_0^{\rm semi} \approx (1+\kappa)^{-1} (\mu_0-\mu_1) (\wh\alpha^{{\rm semi} (t)}-\wh\alpha^{\rm semi} )$,
since the $(2,1)$th component of $\mM^*$ is $(1+\kappa)^{-1} (\mu_0-\mu_1)$.
This suggests that the numerical convergence rate of $\wh\mu_0^{{\rm semi} (t)}$ is mainly determined by that of $\wh\alpha^{{\rm semi} (t)}$.
Since $\wh\alpha^{{\rm semi} (t)}$ converges to $\wh\alpha^{\rm semi} $ linearly, we should have $\wh\mu_0^{{\rm semi} (t)}$ also converges to $\wh\mu_0^{\rm semi} $ linearly.
By similar arguments, we know that $\wh\Sigma_0^{{\rm semi} (t)}$ also converges to $\wh\Sigma_0^{\rm semi} $ linearly.

Unfortunately, the numerical convergence properties of $\wh\mu_1^{{\rm semi} (t)}$ and $\wh\Sigma_1^{{\rm semi} (t)}$ are slightly more complicated.
By Theorem \ref{Theorem 2}, we know that $\wh\mu_1^{{\rm semi} (t+1)}-\wh\mu_1^{\rm semi} \approx - (1+\kappa)^{-1} \Sigma_1 \Sigma_0^{-1} (\wh\mu_0^{{\rm semi} (t)}$
$-\wh\mu_0^{\rm semi} ) + (1+\kappa)^{-1} \Delta_{\mu_1\Sigma_0} \vech(\wh\Sigma_0^{{\rm semi} (t)}-\wh\Sigma_0^{\rm semi}) + (\wh\mu_1^{{\rm semi} (t)}-\wh\mu_1^{\rm semi}) +(1+\kappa)^{-1} \Delta_{\mu_1\Sigma_1} \vech(\wh\Sigma_1^{{\rm semi} (t)}$
$-\wh\Sigma_1)$ and
$\vech(\wh\Sigma_1^{(t+1)}-\wh\Sigma_1) \approx \Delta_{\Sigma_1\mu_0} (\wh\mu_0^{{\rm semi} (t)}-\wh\mu_0^{\rm semi})/(1+\kappa)
+ \Delta_{\Sigma_1\Sigma_0} \vech(\wh\Sigma_0^{^{\rm semi} (t)}-\wh\Sigma_0^{\rm semi} ) /(1+\kappa)
+ \Delta_{\Sigma_1\mu_1} (\wh\mu_1^{{\rm semi} (t)}-\wh\mu_1^{\rm semi} ) /(1+\kappa) - \Delta_{\Sigma_1\Sigma_1} \vech(\wh\Sigma_1^{{\rm semi} (t)}-\wh\Sigma_1^{\rm semi} ) /(1+\kappa)$.
Moreover, the previous discussion suggests that $\wh\mu_0^{{\rm semi} (t)} - \wh\mu_0^{\rm semi} \to 0$ and $\wh\Sigma_0^{{\rm semi} (t)} - \wh\Sigma_0^{\rm semi} \to 0$ as $t\to\infty$.
Thus, for a sufficiently large $t$, we should have
\begin{gather}
%\Delta^{(t+1)} =
\left(\begin{array}{c}
\wh\mu_1^{{\rm semi} (t+1)}- \wh\mu_1^{\rm semi}  \\
\vech(\wh\Sigma_1^{{\rm semi} (t+1)}- \wh\Sigma_1^{\rm semi})  \\
\end{array}\right) \approx
(1+\kappa)^{-1} \left(\begin{array}{cc}
I_p  & \Delta_{\mu_1\Sigma_1} \\
\Delta_{\Sigma_1\mu_1} & -\Delta_{\Sigma_1\Sigma_1} \\
\end{array}\right)
\left(\begin{array}{c}
\wh\mu_1^{{\rm semi} (t)}-\wh\mu_1^{\rm semi}  \\
\vech(\wh\Sigma_1^{{\rm semi} (t)} -\wh\Sigma_1^{\rm semi})  \\
\end{array}\right). \nonumber
\end{gather}
Therefore, whether $\wh\mu_1^{{\rm semi} (t)} \to \wh\mu_1^{\rm semi} $ and $\wh\Sigma_1^{{\rm semi} (t)} \to \wh\Sigma_1^{\rm semi} $ as $t\to\infty$ is mainly determined by the spectral radius of $(I_p, \Delta_{\mu_1\Sigma_1}; \Delta_{\Sigma_1\mu_1}, -\Delta_{\Sigma_1\Sigma_1})/(1+\kappa)$.
According to our numerical analysis, we find that its spectral radius is not always less than 1 unless the fraction number $\kappa$ is sufficiently large.

% \newpage
\section{Numerical Studies}

\subsection{A Simulation Study}

To numerically confirm our theoretical findings, a number of simulation studies are presented.
We independently generate the unlabeled data with sample size $N$ and the labeled data with sample size $m$.
The total sample size is set as $N+m = 10^5$.
The data generation process begins with the unlabeled data.
Specifically, for a given sample $i$ {\color{black}in the unlabeled data}, we generate $Y_i$ independently according to a binomial distribution with $P(Y_i=1)=\alpha$ ($1 \le i \le N$).
To evaluate the rare events effect, a total of five different response probabilities are considered.
They are 50\%, 20\%, 10\%, 1\%, 0.1\%, respectively.
We next generate $X_i$ from a normal distribution $N(-1.5,1)$ if $Y_i=1$.
Otherwise, $X_i$ should be generated from a normal distribution $N(1.5,1)$.
After the unlabeled data are prepared, we independently generate the labeled data $\{(X_i^*,Y_i^*): 1\le\ i \le m\}$ using a similar procedure and their response variables $Y_i^*$s are observed.
To demonstrate the value of partially labeled data, six different percentages of labeled data (i.e., $m/(N+m)\times 100$\%) are evaluated.
They are 0\%, 1\%, 5\%, 10\%, 25\%, 50\%, respectively.
Once the data are generated, the EM and MEM algorithms can be applied to compute the MLE.
The experiment is randomly replicated for $D = 500$ times.

To evaluate the performance of the EM and MEM algorithms, we focus on the numerical convergence performance and the large sample convergence performance.
Numerical convergence relates to the computation cost needed for the initial estimator to numerically converge to the MLE.
This computation cost is reflected by the number of iterations needed to achieve the pre-specified convergence criterion, and is fundamentally determined by the spectral radius of the contraction operator evaluated at the true value.
Therefore, both $N_{\rm iter}$ and $\rho\{ \dot\mF(\theta) \}$ serve as useful measures of numerical convergence.
Specifically, the number of iterations is denoted as $N_{\rm iter}^{(d)}$ in the $d$th replication.
We next define $N_{\rm iter} = D^{-1}\sum_{d=1}^D N_{\rm iter}^{(d)}$ as an overall measure.
Denote the spectral radius of the contraction operator $\dot\mF(\theta)$ as $\rho\{ \dot\mF(\theta) \}^{(d)}$ in the $d$th replication.
We next define $\rho\{ \dot\mF(\theta) \} = D^{-1}\sum_{d=1}^D \rho\{ \dot\mF(\theta) \}^{(d)}$ as an overall measure.
In contrast, large sample convergence refers to the discrepancy between the MLE and the true parameter.
To evaluate the estimation accuracy, we calculate the Root Mean Square Error (RMSE) as $\mbox{RMSE}=q^{-1}\sum_{j=1}^{q} \big\{D^{-1} \sum_{d=1}^{D}(\wh{\theta}_j^{(d)} -\theta_j)^2\big\}^{1/2}$, where $\wh{\theta}^{(d)}$ is the MLE computed in the $d$-th random replication.
The detailed results are summarized in Table \ref{t: 2pp}.

\begin{table}[htbp]
\centering
\caption{
Detailed simulation results with different response probabilities and labeling percentages.
The estimated RMSE, the averaged number of iterations $N_{\rm iter}$ and the averaged spectral radius of the contraction operator $\rho\{ \dot\mF(\theta) \}$ are reported.
}
\begin{tabular}{ccccccccc}
\hline\hline
\multicolumn{2}{c}{$m/(m+N) \times 100\%$}  & 0\%     & 1\%  & 5\%  & 10\%   & 25\%  & 50\% \\
\hline % Percentage of Positive Instances Percentage of Labeled Data
\multirow{3}[2]{*}{$\alpha=50\%$} & RMSE  & 0.0092 & 0.0087 & 0.0074 & 0.0067 & 0.0057 & 0.0052 \\
& $\rho\{ \dot\mF(\theta) \}$ & 0.9323 & 0.9229 & 0.8856 & 0.8389 & 0.6990 & 0.4661 \\
& $N_{\rm iter}$ & 119.64 & 106.42 & 74.14 & 53.68 & 28.91 & 15.00 \\
\hline
\multirow{3}[2]{*}{$\alpha=20\%$} & RMSE  & 0.0120 & 0.0113 & 0.0095 & 0.0078 & 0.0068 & 0.0061 \\
& $\rho\{ \dot\mF(\theta) \}$ & 0.9419 & 0.9324 & 0.8949 & 0.8479 & 0.7065 & 0.4710  \\
& $N_{\rm iter}$
& 146.69 & 127.52 & 84.37 & 59.34 & 30.86 & 16.00  \\
\hline
\multirow{3}[2]{*}{$\alpha=10\%$} & RMSE  & 0.0165 & 0.0151 & 0.0126 & 0.0102 & 0.0084 & 0.0074 \\
& $\rho\{ \dot\mF(\theta) \}$ & 0.9533 & 0.9437 & 0.9057 & 0.8579 & 0.7150 & 0.4768  \\
& $N_{\rm iter}$ & 176.60 & 148.69 & 92.02 & 62.54 & 31.37 & 16.00 \\
\hline
\multirow{3}[2]{*}{$\alpha=1\%$} & RMSE  & 0.0663 & 0.0529 & 0.0393 & 0.0306 & 0.0245 & 0.0199 \\
& $\rho\{ \dot\mF(\theta) \}$ & 0.9839 & 0.9735 & 0.9344 & 0.8852 & 0.7382 & 0.4919 \\
& $N_{\rm iter}$ & 436.16 & 274.70 & 122.57 & 75.00 & 33.92 & 16.29 \\
\hline
\multirow{3}[2]{*}{$\alpha=0.1\%$} & RMSE  & 0.2668 & 0.1932 & 0.1363 & 0.1055 & 0.0725 & 0.0593 \\
& $\rho\{ \dot\mF(\theta) \}$ & 0.9992 & 0.9889 & 0.9488 & 0.8989 & 0.7493 & 0.5019 \\
& $N_{\rm iter}$ & 733.78 & 456.11 & 147.65 & 82.34 & 34.55 & 16.39 \\
\hline\hline
\end{tabular}%
\label{t: 2pp}%
\end{table}%

From Table \ref{t: 2pp}, we can draw the following conclusions.
First, we find that for a fixed percentage of the labeled data, the RMSE value increases as the response probability $\alpha$ decreases.
This indicates that the statistical efficiency of the MLE deteriorates as rare events become rarer (i.e., $\alpha \to 0$).
For example, when $m/(m+N)=0\%$, the RMSE value of $\wh\theta$ is 0.0663 with $\alpha=1\%$, which is much larger than 0.0092 of the case with $\alpha=50\%$.
Second, we observe that for a fixed percentage of the labeled data, the spectral radius $\rho\{ \dot\mF(\theta) \}$ increases as the response probability $\alpha$ decreases.
Consider for example the case with $\alpha=1\%$ and $m/(m+N)=0\%$.
In this case, we find $\rho\{ \dot\mF(\theta) \} =0.9839$, which is very close to 1.
This explains why the numerical convergence of a standard EM algorithm could be extremely slow.
The resulting $N_{\rm iter}$ value could be as large as $N_{\rm iter} =436.16$.
Third, for a fixed response probability, we find that the RMSE value decreases as the percentage of labeled data increases.
This demonstrates that a higher percentage of the labeled data would significantly improve the statistical efficiency.
For example, when $\alpha=1\%$, the RMSE value of $\wh\theta$ is 0.0663 with $m/(m+N)=0\%$, which is much larger than the 0.0306 of the case with $m/(m+N)=10\%$.
Lastly, we observe that for a fixed response probability, both the $\rho\{ \dot\mF(\theta) \}$ and $N_{\rm iter}$ values decrease as the percentage of the labeled data increases.
Consider for example the case with $\alpha=10\%$.
When $m/(m+N)=1\%$, we have $\rho\{ \dot\mF(\theta) \} =0.9437$ and $N_{\rm iter} =148.69$.
As $m/(m+N)=10\%$, we have $\rho\{ \dot\mF(\theta) \} =0.8579$ and $N_{\rm iter} =62.54$.
This suggests that a slightly enhanced label percentage could lead to much improved numerical convergence performance for the proposed MEM algorithm.
All these observations are in line with the theoretical findings of Theorems 1 and 2.

\subsection{A Real Data Example}

\begin{figure}[t!]
\centering
\includegraphics[width=0.6 \textwidth]{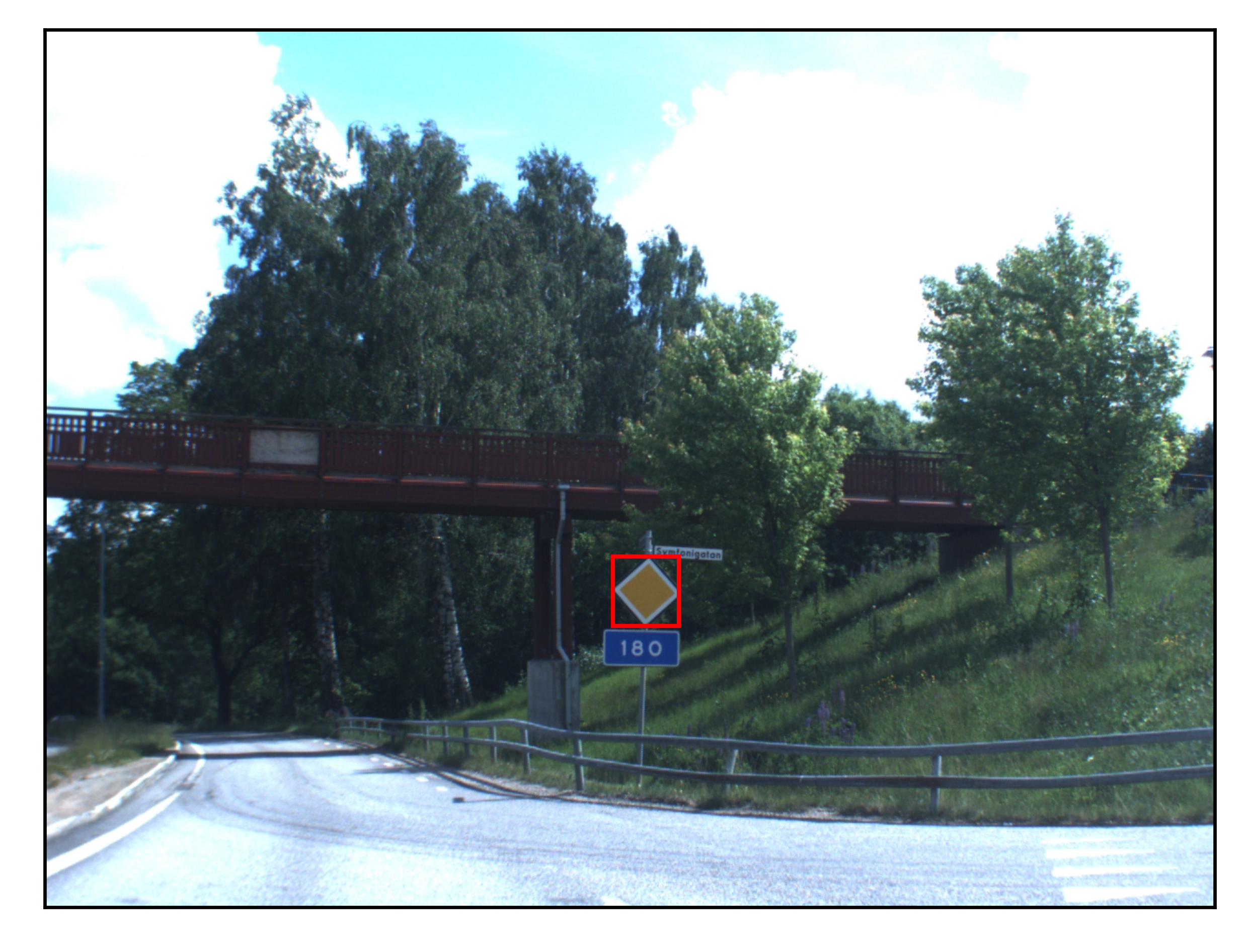}\par
\caption{
An example from the STS dataset for traffic sign detection.
The original image is of size $960 \times 1,280 \times 3$.
%The red bounding box is used to annotate a local region containing a traffic sign.
The red bounding box of a small object treated as a positive instance only covers less than 1\% of the original image.}
\label{f: bbox}
\end{figure}

To demonstrate the practicality of our proposed algorithm,
we present an interesting real data example for traffic sign recognition.
The dataset used here is the Swedish Traffic Signs (STS) dataset \citep{10.1007/978-3-642-21227-7_23,larsson2011correlating}, which can be publicly obtained from \textit{https://www.cvl.isy.liu.se/research/datasets/traffic-signs-dataset/}.
The dataset contains a total of 1,970 high--resolution (960 $\times$ 1,280) color images with annotation.
A graphical illustration of the labeled data can be found in Figure \ref{f: bbox}.
The objective here is to automatically detect the traffic signs in Figure \ref{f: bbox}.
To this end, we randomly partition all labeled data into two parts.
The first part contains about 80\% of the whole data for training. The remaining 20\% part is used for testing.
Subsequently, we demonstrate how this task can be converted into a problem, which can be efficiently solved by our proposed method.

We start with the construction of the feature vector.
Specifically, for each original high--resolution image of $960 \times 1,280$, a pretrained VGG16 model is applied \citep{simonyan2014very}.
This results in a feature map of size $30 \times 40 \times 512$.
This feature map can be viewed as a new ``image" with resolution $30 \times 40$ and a total of 512 channels.
In this regard, each pixel in this feature map can be treated as one sample with a feature vector of $p=512$ dimensions.
This leads to the feature vectors as $\{ \mX_{i,k_1,k_2}\in \mR^{p}: 1\le k_1\le30, 1\le k_2\le40\}$.
Therefore, a total of $30 \times 40 =1,200$ samples can be obtained for each image.
Because a total of 1,970 images are involved in the whole dataset, the total sample size in this study is then $N = 1,970 \times 1,200 = 2,364,000$.
Next, for each $\mX_{i,k_1k_2}$, we generate a binary response $\mY_{i,k_1,k_2}$ based on whether a traffic sign is involved in the pixel location.
This leads to labels as $\{ \mY_{i,k_1,k_2} \in \mR: 1\le k_1\le30, 1\le k_2\le40\}$.
The number of positive instances accounts for only 0.225\% of the total sample, which is expected because the region containing traffic signs in a high--resolution image is extremely small.
Once the data are well prepared, the training data are randomly split into two parts.
In the first part, the labels of the samples are treated as if they were missing.
The second part is treated as if the labels of the samples were observed.
Thereafter, the MEM algorithm can be applied to the training data.
For a comprehensive evaluation, six different labeled percentages (i.e., $m/(N+m)\times 100$\%) are evaluated.
They are, respectively, 0\%, 5\%, 25\%, 50\%, 75\% and 100\%.
For a reliable evaluation, the experiment is randomly replicated for $D=20$ times.

\begin{figure}[htbp]
\centering
\includegraphics[width=1\textwidth]{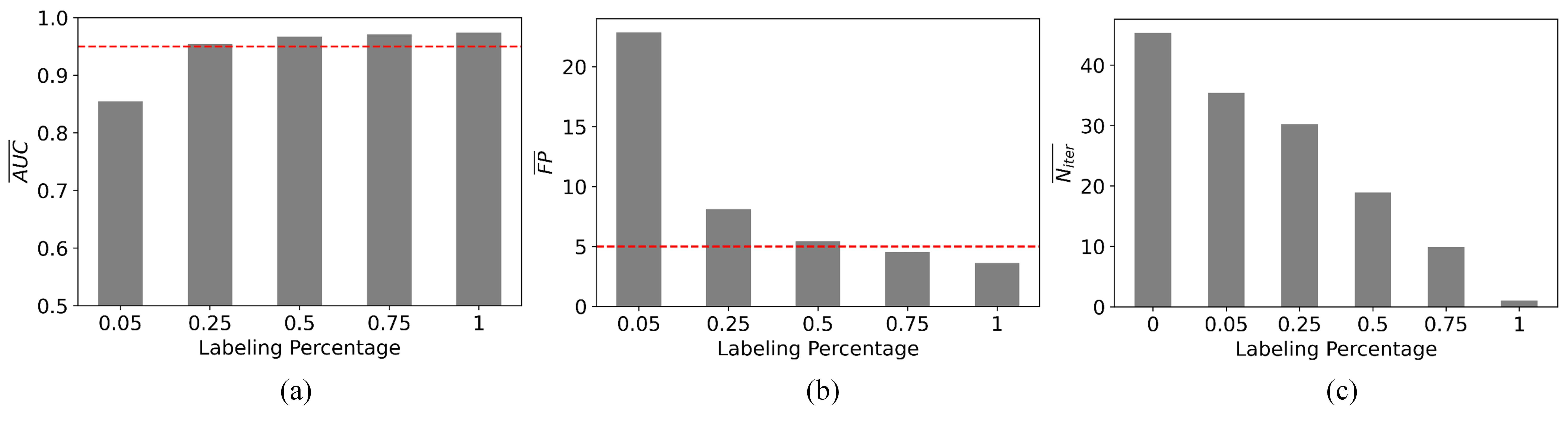}\par
\caption{Detailed prediction results obtained from the test data.
The left panel demonstrates the $\overline{\rm AUC}$ results.
The red dashed line corresponds to $\overline{\rm AUC}=0.95$.
The middle panel presents the $\overline{\rm FP}$ results.
The red dashed line corresponds to $\overline{\rm FP}=5$.
The right panel shows the $\overline{N_{\rm iter}}$ results.
%	The bottom  panel shows the $\overline{Time}$ results.
}
\label{f: preds}
\end{figure}

To gauge the finite sample performance of the proposed method, three different measures are used.
The first measure is area under the curve (AUC) \citep{ling2003auc}.
Consider the $i^*$th image ($1\le i^* \le N^*$) in the test data, where $N^*=394$ denotes the number of images for testing.
For a given pixel $(k_1,k_2)$ and one particular estimator $\wh\theta$ obtained from the train data, we then estimate the response probability of $\wh{\pi}_{i^*,k_1,k_2}= f_{\wh\theta}(\mX_{i^*,k_1,k_2})^{-1} \alpha \phi_{\wh\mu_1,\wh\Sigma_1}(\mX_{i^*,k_1,k_2})$.
Accordingly, the AUC measure can be computed at the image level and denoted by ${\rm AUC}_{i^*}$.
The image-level AUC is then averaged across different images, and denoted by AUC$^* = N^{*-1} \sum_{i^*} {\rm AUC}_{i^*}$.
The replicated AUC$^*$ values are averaged across the random replications and are denoted as $\overline{\rm AUC}$.
The second measure is the number of false positives.
For a given random replication and the $i^*$th image, we can predict $\wh\mY_{i^*,k_1,k_2} = I\big(\wh{\pi}_{i^*,k_1,k_2}>c_{i^*}\big)$,
where $c_i^*$ is the largest threshold value so that all positive instances can be correctly captured.
However, the price is inevitably paid by mistakenly predicting some negative samples as positives.
We define the number of false positives for the $i^*$th image in the test data as ${\rm FP}_{i^*} = \sum_{k_1,k_2} I\big(\wh{\mY}_{i^*,k_1,k_2}=1 \big) I\big(\mY_{i^*,k_1,k_2}=0 \big)$.
Its median value is then computed as ${\rm FP^*}$.
Then its overall mean across different random replications is denoted by $\overline{\rm FP}$.
The third measure is the number of iterations used for the numerical convergence and denoted as $N_{\rm iter}^*$.
The averaged $N_{\rm iter}^*$ values across the random replications are computed and denoted as $\overline{N_{\rm iter}}$ as an overall measure.
The prediction results are presented in Figure \ref{f: preds}.

We find that the $\overline{\rm AUC}$ value is 0.5235 and $\overline{\rm FP}$ value is 1,098.85 when the labeling percentage is 0\%.
These are totally not comparable with the other cases and thus not presented in Figure \ref{f: preds}.
This suggests that the prediction performance of the standard EM algorithm for a GMM with rare events could be extremely poor.
However, the performance can be significantly improved if the data are partially labeled.
A quick glance at Figure \ref{f: preds} suggests that cases with labeling percentages larger than 25\% perform well in terms of $\overline{\rm AUC}$ values.
For example, consider the case with a labeling percentage being 25\%. The $\overline{\rm AUC}$ value becomes 0.9547, which is slightly worse than the 0.9711 for the fully labeled case.
Similar observations are also observed for other performance measures (i.e., $\overline{\rm FP}$ and $\overline{N_{\rm iter}}$).
Specifically, when the labeling percentage is 50\%, the $\overline{\rm FP}$ value becomes 5.45, which is slightly worse than the 3.63 value obtained in the fully labeled case.
Meanwhile, the $\overline{N_{\rm iter}}$ value becomes 18.95, which performs much better than the 45.35 value obtained in the fully unlabeled case.

\section{Concluding Remarks}

In this study, we focus on a problem related to the analysis of rare events data.
Statistical analysis of rare events data differs significantly from that of regular data, and considerable progress has been made in the field concerning rare events data analysis \citep{nguyen2012comparative,wang2020logistic,wang2021nonuniform,triguero2015evolutionary,triguero2016evolutionary,duan2020self}.
However, the existing methods often suffer from one common limitation.
That is the dataset must be fully labeled.
Thus, the problem of unsupervised or semi-supervised learning remains open for discussion.
In our first attempt, we investigate here a GMM with rare events.
The EM algorithm has been commonly used to estimate GMM \citep{wu1983convergence,xu1996convergence}.
%Its theoretical properties have been studied in the past literature.
%However, no rigorously asymptotic theory has been developed in this regard.
Consequently, we study the theoretical properties of the standard EM algorithm for a GMM with rare events.
%In this study, we investigate the numerical convergence properties of the standard EM algorithm for a GMM with rare events.
To this end, we formulate an EM algorithm for a GMM with rare events as an iterated algorithm governed by a contraction operator.
Our results suggest that the numerical convergence rate of the standard EM algorithm is extremely slow.
To address this, we develop an MEM algorithm.
% with a partially labeled dataset
We theoretically demonstrate that the numerical convergence rate of this algorithm is significantly higher than that of a standard EM algorithm if the labeled percentage is carefully designed.

To conclude this work, we would like to discuss some intriguing avenues for future research.
Firstly, in this study, our focus was solely on the GMM as a parametric model. However, given the widespread use of various learning methods, such as deep neural networks, it would be of great interest to explore more complex and general models for rare events data in future research projects.
Secondly, although we examined a GMM with only two underlying classes, our theoretical results can be extended to scenarios involving multiple classes with minimal modifications. Nevertheless, contemporary classification problems often involve a growing number of classes.
Exploring how to extend our theoretical findings to address such challenging cases would be an intriguing direction for future exploration.
Lastly, most existing literature on semi-supervised learning, including this study, has assumed that both positive and negative instances in previously labeled samples are accurately annotated.
However, in situations where positive instances are rare, we often encounter cases where only the labels for positive instances are practically available, whereas the labels for negative instances are too numerous to be analytically provided. This presents an interesting scenario where only positive instances are practically labeled. Exploring solutions for this problem is another fascinating topic worthy of investigation.

% Acknowledgements and Disclosure of Funding should go at the end, before appendices and references

\acks{Jing Zhou's research is supported in part by the National Natural Science Foundation of China (No.72171226, 11971504) and the National Statistical Science Research
Project (No.2023LD008).
Hansheng Wang's research is partially supported by the National Natural Science Foundation of China (No.12271012)}


\vskip 0.2in
%\bibliography{sample}
\bibliography{reference}

\end{document}